\documentclass[a4paper,11pt]{article}
\pdfoutput=1 

\usepackage{jcappub} 

\usepackage[T1]{fontenc} 
\usepackage{subfig}
\usepackage{lscape}
\usepackage{cleveref}
\usepackage{amssymb}

\usepackage[figuresright]{rotating}

\usepackage{graphicx}
\usepackage{xspace}
\usepackage{units}
\usepackage{xcolor}
\usepackage{subfig}
\usepackage{multirow}
\usepackage{cleveref}
\def\Xmax{\ifmmode {X_\mathrm{max}}\else{$X_\mathrm{max}$}\fi\xspace}
\def\sigmaXmax{\ifmmode {\mathrm{RMS}(X_\mathrm{max})}\else{RMS$(X_\mathrm{max})$}\fi\xspace}
\def\meanXmax{\ifmmode {\langle X_\mathrm{max}\rangle}\else{$\langle X_\mathrm{max}\rangle$}\fi\xspace}
\def\eV{\ifmmode {\mathrm{\ e\kern -0.1em V}}\else
\textrm{e\kern -0.1em V}\fi\xspace}
\def\gcm{\ifmmode {\mathrm{g/cm}^2}\else{g/cm$^2$}\fi\xspace}
\newcommand{\energy}[1]{\unit[$10^{#1}$]{\eV}}

\newcommand{\bea}{\begin{array}}
\newcommand{\ear}{\end{array}}

\newcommand{\bege}{\begin{equation}}
\newcommand{\enge}{\end{equation}}

\newcommand{\beq}{\begin{eqnarray}}\newcommand{\benu}{\begin{enumerate}}\newcommand{\enu}{\end{enumerate}}
\newcommand{\eeq}{\end{eqnarray}}

\author[a,b,1]{Rita C. dos Anjos,\note{Corresponding author.}}
\author[b]{Vitor de Souza}
\author[c]{Rogerio M. de Almeida}
\author[d]{Edivaldo M. Santos}

\affiliation[a]{Departamento de Engenharias e Exatas, Universidade Federal do Paran\'a (UFPR),\\
Pioneiro, 2153, 85950-000 Palotina, PR, Brazil.}
\affiliation[b]{Instituto de F\'isica de S\~ao Carlos, Universidade de S\~ao Paulo, CP 369, 13560-970, S\~ao Carlos, SP, Brasil.}
\affiliation[c]{EEIMVR, Universidade Federal Fluminense, Volta Redonda, RJ, Brazil.}
\affiliation[d]{Instituto de F\'{\i}sica, Universidade de S\~ao
  Paulo, Rua do Mat\~ao trav. R 187, 05508-090, S\~ao Paulo, Brazil.}

\emailAdd{ritacassia@ufpr.br}
\emailAdd{vitor@ifsc.usp.br}
\emailAdd{rmenezes@id.uff.br}
\emailAdd{emoura@if.usp.br}

\title{The influence of the observatory latitude on the study of ultra high energy cosmic rays}

\abstract{Recent precision measurements of the Ultra High Energy Cosmic Rays (UHECR) arrival directions, spectrum and parameters related to the mass of the primary particle have been done by the HiRes, Pierre Auger and Telescope Array (TA) Observatories. In this paper, distributions of arrival directions of events in the nearby Universe are assumed to correlate with sources in the 2MASS Redshift Survey (2MRS), IRAS 1.2 Jy Survey, Palermo Swift-BAT and Swift-BAT catalogs, and the effect of the latitude of the observatory on the measurement of the energy spectrum and on the capability of measuring anisotropy is studied. The differences between given latitudes on the northern and southern hemispheres are quantified. It is shown that the latitude of the observatory: a) has an influence on the total flux measured and b) imposes an important limitation on the capability of measuring an anisotropic sky.}

\begin{document}
\maketitle
\flushbottom


\section{Introduction}

The international community studying UHECR has recently done a big effort in the direction of a common interpretation of the data measured by several observatories~\cite{bib:uhecr2010,bib:uhecr2014}. However, bringing together measurements performed by different experiments is not an easy task. In the past, the discrepancies between the measurements from several observatories have been ascribed to experimental particularities and analysis procedures. Detector uncertainties and biases, different analysis assumptions, lack of events or discrepant interpretations based on different extrapolations of the hadronic interactions properties have been enough to explain all the differences among the measurements. Presently, the improvement of our understanding about the detection techniques, the multiplication of significant parameters extracted from the air shower, the construction of large and stable observatories and the extension of accelerator data to even higher energies have minimized the unknown contributions to the differences between the most important quantities measured by several observatories.

In this new era, precision measurements of the UHECR energy spectrum, arrival directions and parameters related to the mass of the primary particle have been done by the HiRes~\cite{bib:hires:spectrum}, the Pierre Auger and TA Observatories~\cite{bib:auger:spectrum,bib:ta:spectrum}. The Pierre Auger Observatory has measured a correlation of events with energy above 57 EeV with AGNs closer than 75 Mpc~\cite{bib:auger:science}. In the following years, the strength of the correlation has decreased with the accumulation of larger statistics and the most updated results are compatible with isotropy of events with energy above 53 EeV~\cite{bib:auger:anisotropy} at the 2$\sigma$ level. On the other hand, the HiRes Collaboration has published a similar study with less statistics in which no correlation with AGNs is seen~\cite{bib:hires:anisotropy} and recently the TA Collaboration also reported no statistically significant correlation with AGNs~\cite{bib:ta:anisotropy}. Despite the lack of correlation with AGNs, TA has recently reported indications of anisotropy centered at R.A. = 146$^\circ$7, decl. = 43$^\circ$2 with 20$^\circ$ scale~\cite{bib:ta:hotspot}.

The energy spectrum measured by the HiRes~\cite{bib:hires:spectrum}, the Pierre Auger~\cite{bib:auger:spectrum} and TA~\cite{bib:ta:spectrum} Observatories agree remarkably well in the shape up to $10^{19.6}$ eV, but show an offset in the total measured absolute flux. It has been shown that the differences in the total measured flux can be explained by energy shifts within the estimated systematic uncertainty in the reconstructed energy of each experiment~\cite{bib:cern:spectrum}.

The most reliable technique used by these observatories to determine the UHECR composition is the measurement of the atmospheric depth at which the shower reaches its maximum (\Xmax), as determined by telescopes that detect the fluorescence light emitted by air molecules. The comparison of the data measured by the experiments is not straightforward due to different analyses used by each group~\cite{bib:cern:composition}. The Pierre Auger Observatory measured a significant change in the trend of \meanXmax with energy at \energy{18.27}~\cite{bib:auger:er,bib:auger:er:long}, which can be interpreted as an increase of the abundance of heavier elements in the measured data~\cite{bib:auger:xmax:interpretation}. The \Xmax data measured by the HiRes and TA Observatories are statistically compatible with constant abundance, but also with a changing composition (as suggested by Auger) in the energy range from \energy{18} to \energy{19}~\cite{bib:cern:composition,bib:hires:er,bib:ta:er}.

The comparison of the main measurements made by the three observatories shows discrepancies that could go beyond the detection particularities and differences in the analyses. If the extragalactic magnetic field is not extreme and the UHECR particles are not all heavy, the deviation of the highest energy particles ($E > 10^{19.6}$ \eV) is not expected to be large~\cite{bib:globus}. Under these assumptions, it is expected that the measurement depends on the location of the observatory on Earth. It has been already shown that the normalization of the flux measured by each experiment might depend on the latitude of the Observatory~\cite{bib:glu}.

The calculation presented in this paper quantifies the differences in
flux at a few latitudes on Earth for $E > 10^{19.6}$ \eV. The effect
of the latitude on the capability of an observatory to determine an
anisotropy signal is also investigated. The catalogs 2MASS Redshift
Survey (2MRS)~\cite{bib:2mass}, IRAS 1.2 Jy Survey~\cite{bib:survey},
Palermo Swift-BAT~\cite{bib:palermo} and Swift-BAT~\cite{bib:bat} have
been used as templates for the UHECR sources distribution. Calculations
were done with the incomplete catalogs, as they are published, but we
have also completed the original catalogs with sources isotropically
distributed in the sky and whose distances are such that the number of
sources in small bins of redshift scales as $\propto z^2$. The
incompleteness of the catalogs is discussed along the paper when
necessary. Throughout the paper sources have been considered to have equal UHECR luminosity. The contribution of each source to the flux measured by each observatory is calculated taking into account its exposure function.

The organization of the paper is as follows. In Section~\ref{sec:spectrum}, the influence of the latitude of the experiment on the energy spectrum measurement is evaluated. In Section~\ref{sec:anisotropy}, the influence of the latitude on the capability of measuring an anisotropy signal is shown. In Section~\ref{sec:conclusion}, the conclusions are presented.

\section{Effect of the latitude on the measured energy spectrum}
\label{sec:spectrum}

The contribution of a point source to the total flux measured on Earth can be written as:
\begin{equation}
J_{CR}^s(E) =  \frac{W_s}{4 \pi D_s^2 (1+z)} \Phi_0(E),
\label{eq:jcr3}
\end{equation}
where $\Phi_o(E)$ is the energy spectrum at the source, $D_s$ is the
comoving distance of the source from Earth, $z$ the redshift of the
source and $W_s$ is the exposure. For a full detection efficiency, the
exposure can be calculated analytically~\cite{bib:sommers}. In order to taking into account the
deflections due to magnetic fields along the particle propagation, we
have performed a Gaussian smearing with 30 degrees of resolution on the
resulting coverage map using the \emph{Coverage and Anisotropy Toolkit}~\cite{bib:toolkit} developed by members of the Pierre Auger Collaboration as a tool to consider the deflections due to extragalactic magnetic field in the construction of sky maps. We choose a large angle of 30 degrees to show that the difference in the measured energy spectrum between hemispheres is significant even for large values of the smearing angle.

In order to study the effect of the latitude on the measured energy spectrum, the relative contribution of the sources in the nearby Universe ($z < 0.072$) in the 2MASS Redshift Survey (2MRS), IRAS 1.2 Jy Survey, Palermo Swift-BAT and Swift-BAT catalogs was calculated. The contribution of each source received a weight ($P_s$) given by
\begin{equation}
P_s = \frac{W_s}{4 \pi D_s^2 (1+z)}.
\label{eq:peso}
\end{equation}
The weight was calculated for observatories placed at four latitudes in each of the northern and southern hemispheres ($\pm$ 35 and $\pm$ 55 degrees). Sources have been divided in shells of 10 Mpc distance from Earth.

\begin{figure}[!h]
  \begin{center}
    \includegraphics[angle=90,width=0.49\textwidth]{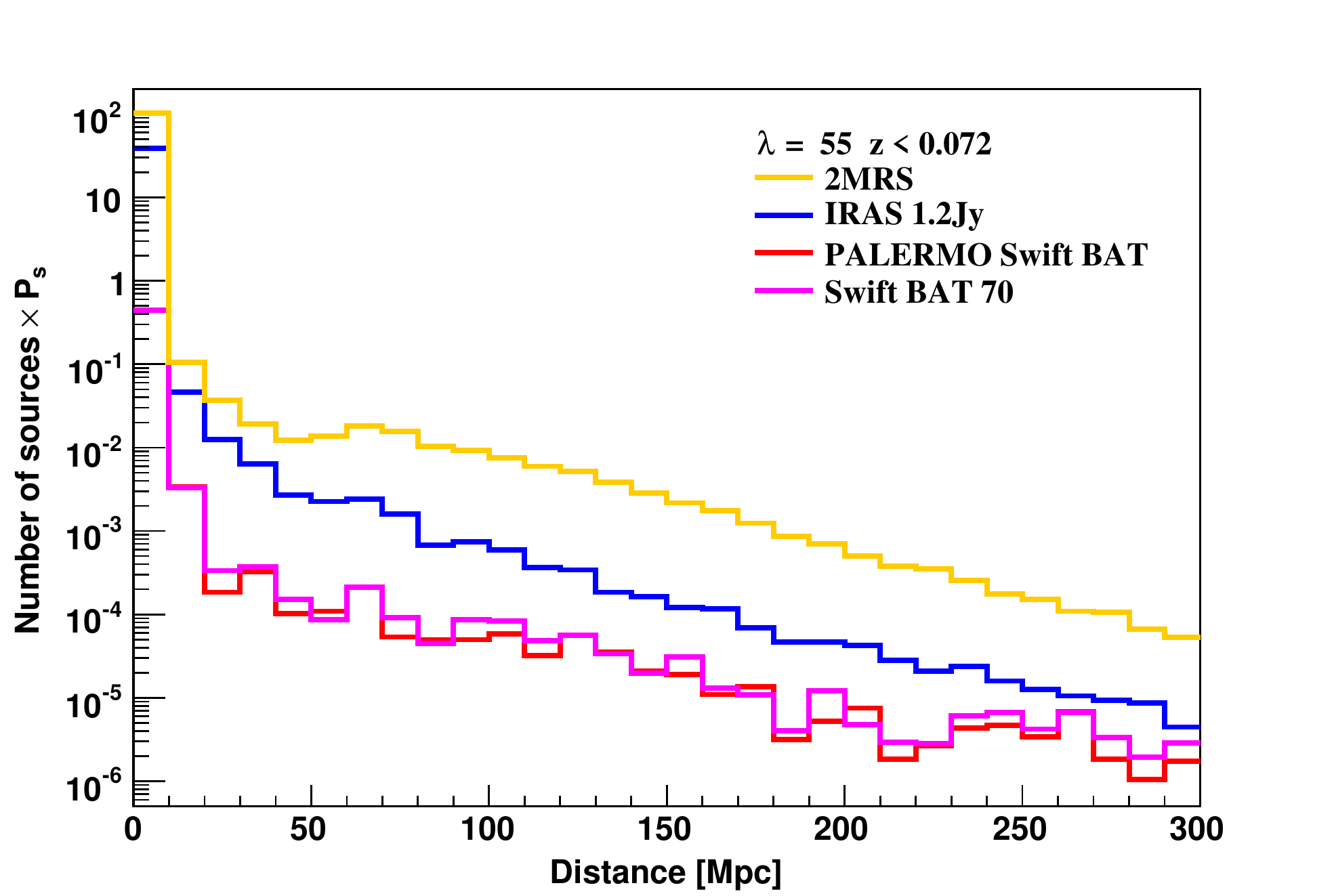}
    \includegraphics[angle=90,width=0.49\textwidth]{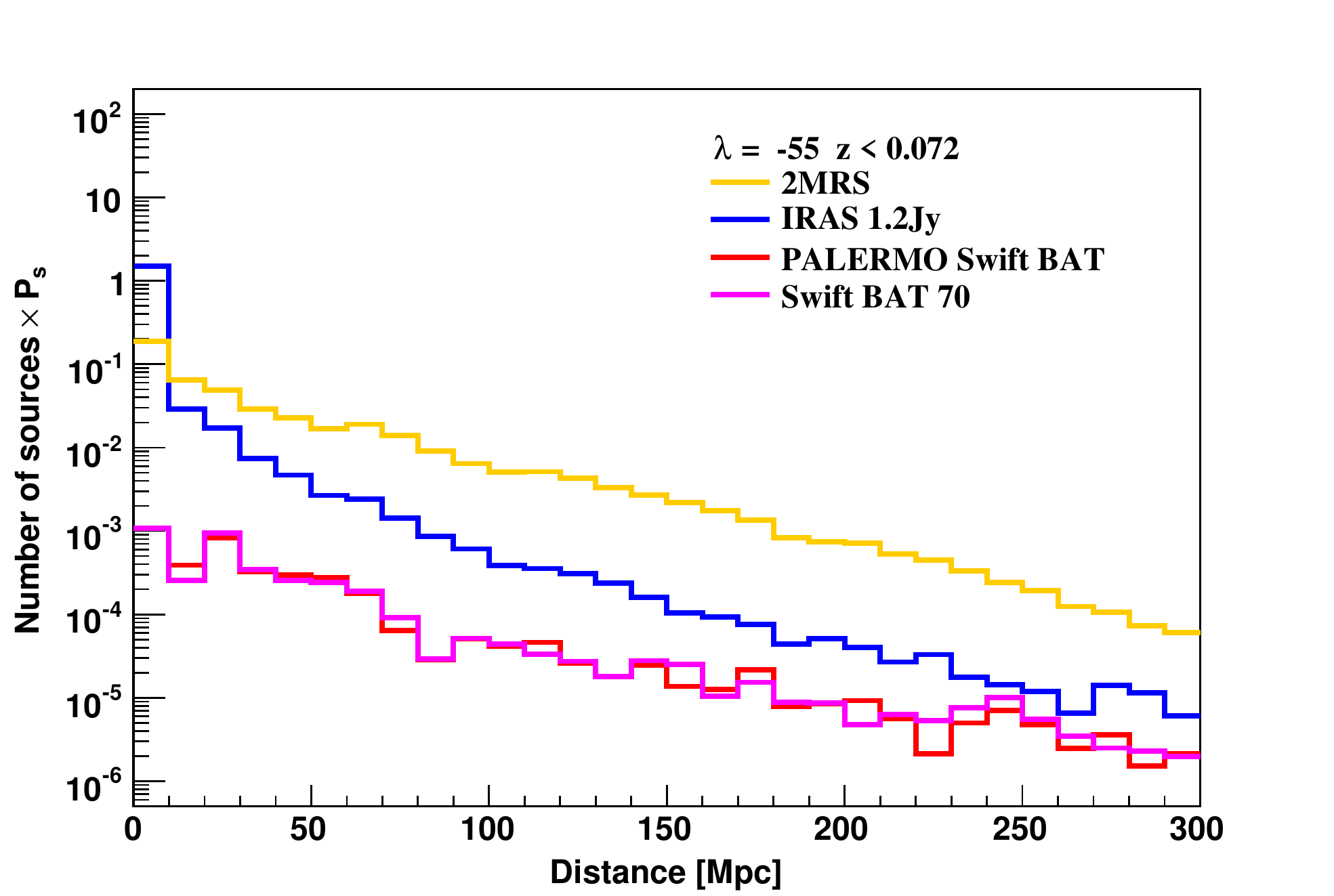}
    \caption{Histograms of number of sources multiplied by the exposure versus distance from Earth. The exposure has been calculated for observatories located at $\pm$ 55 degrees of latitude. The left figure shows the northern locations and the right figure shows the southern locations. The relative contribution to the flux from each bin in distance considered is shown. It is also possible to note differences between southern and northern latitudes.}
    \label{fig:weight:norte:sul}
  \end{center}
\end{figure}

\begin{figure}[!h]
  \begin{center}
    \includegraphics[angle=90,width=0.49\textwidth]{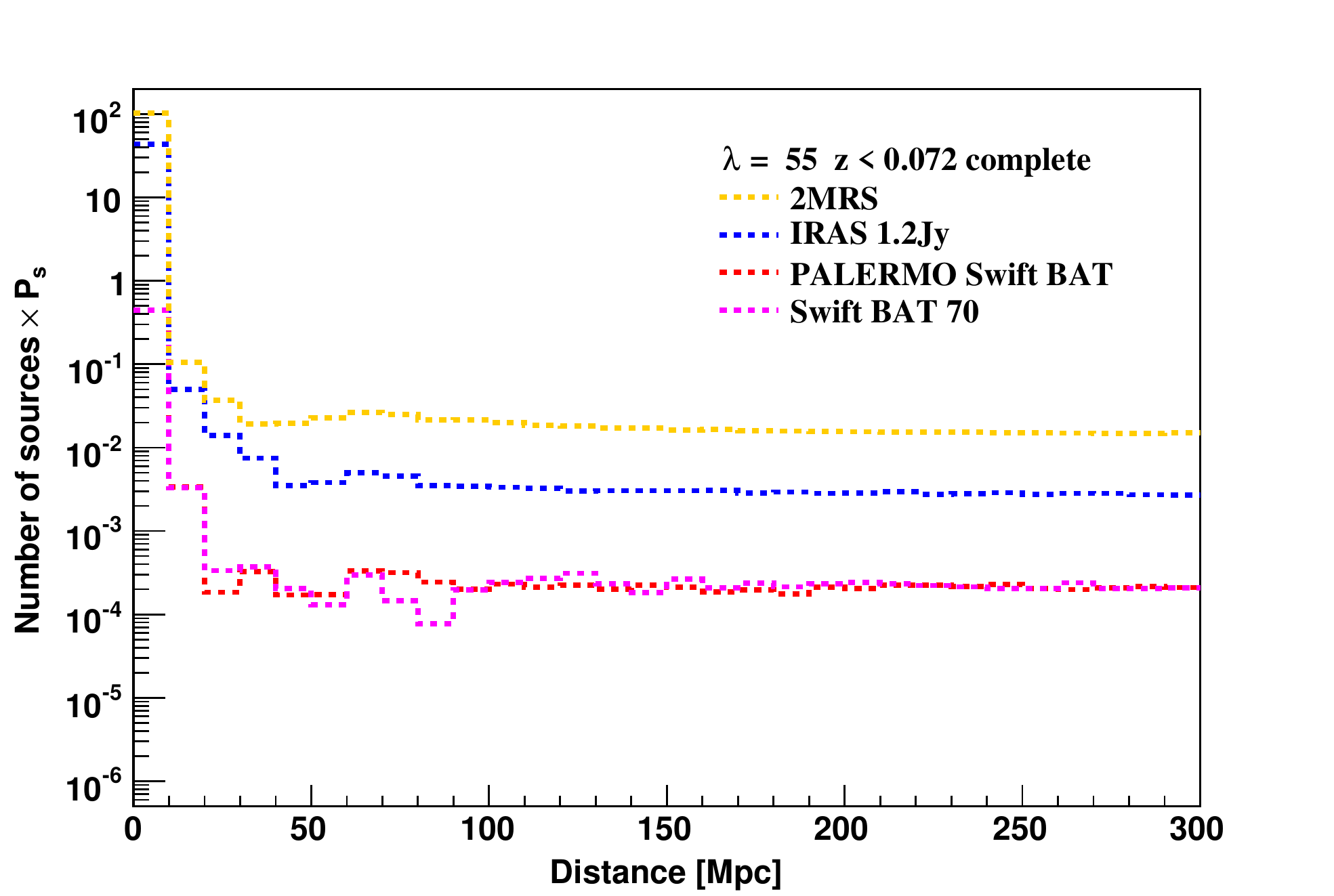}
    \includegraphics[angle=90,width=0.49\textwidth]{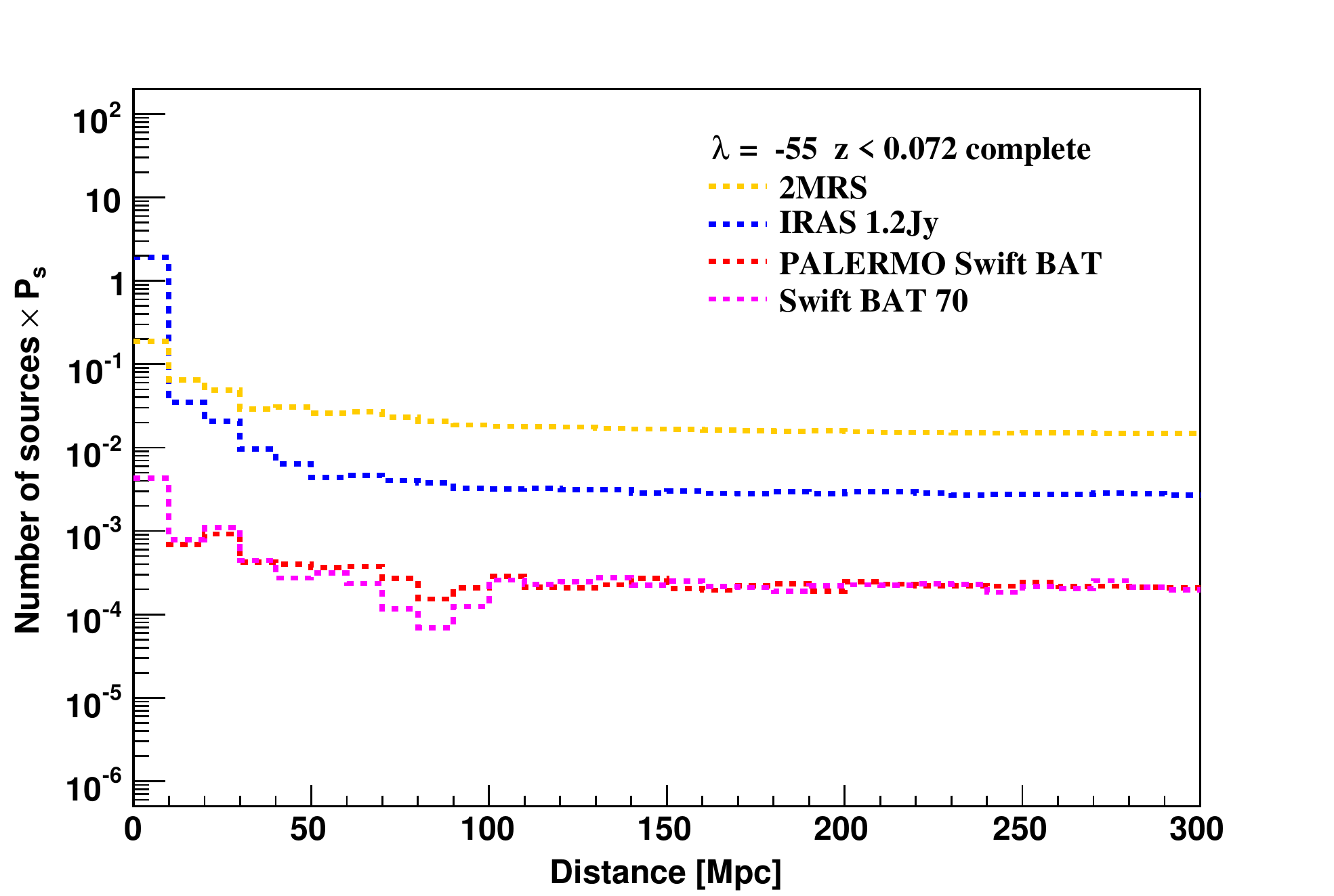}
    \caption{Histograms of number of sources to completed catalogs multiplied by the exposure versus distance from Earth. The exposure has been calculated for observatories located at $\pm$ 55 degrees of latitude. The left figure shows the northern locations and the right figure shows the southern locations. The relative contribution to the flux from each bin in distance considered is shown. It is also possible to note differences between southern and northern latitudes.}
    \label{fig:weight:norte:sul:complete}
  \end{center}
\end{figure}

Figures~\ref{fig:weight:norte:sul}
and~\ref{fig:weight:norte:sul:complete} show the relative contribution
of each distance bin to the total flux measured on Earth at latitudes $\pm$ 55
degree. Figure~\ref{fig:weight:norte:sul} was calculated with the
sources in each catalog as published. In
Figure~\ref{fig:weight:norte:sul:complete}, we have completed the
original catalog with sources isotropically distributed in the sky, so that
the number of sources scales as $\propto z^2$ in bins of redshift. The
dependence on the latitude is clear in both figures: I) the northern
hemisphere is exposed to a larger flux than the southern hemisphere and
II) the relative contribution of each shell in distance is different for
each latitude. If we consider the catalogs completed, the effect is
pronounced only up to 70 Mpc. This in turn could imply a difference in
the composition determined by each observatory since the abundance is
affected by the propagation of the particles in the intergalactic
medium.

Figures~\ref{fig:weight:norte:sul} and~\ref{fig:weight:norte:sul:complete} also show that the differences between the Northern and Southern skies are dominated by local sources ($D_s < 100$ Mpc). The further the source, less it contributes to the total measured flux and more isotropic the sky is. The ratio between Northern and Southern hemiphere of the integral of the quantity (Number of sources $\times P_s$) shown in~\ref{fig:weight:norte:sul:complete} is basically one for distance larger than 100 Mpc. The incompletness of the the catalogs regarding the obscurance due to the galactic plane is also neglegible for the studies presented in this paper. The solid angle covered by the galactic plane for observers in the Northern and Southern hemispheres are identical and correspods to 8.6\% of each coverage. The conclusion below are based on differences of fluxes measured by observatories in Northern and Southern hemispheres. Therefore the results presentd here are valid under the assumption that the obscured sky seen by a Northern and Southern observatory have the same isotropic distribution of sources.

The difference between the energy spectrum measured by northern and
southern observatories for energies above $10^{19.6}$ eV was studied in
detail including the propagation of the particles in the intergalactic
medium. We pursue the analysis of the specific astrophysical scenario
with sources in the nearby Universe, since the fraction of surviving
hadrons with $E > 10^{19.6}$ eV is relevant up to $z\sim
0.072$~\cite{bib:kalashev}. For each source in the catalogs with $z <
0.072$, 50,000 events have been generated and propagated through the
intergalactic medium to Earth. The 1D propagation was done using the
CRPropa (v. 2.0) program~\cite{bib:crpropa}. \mbox{CRPropa} is a public
software to simulate the propagation of nuclei in the intergalactic
medium, taking into account the most important interactions and radiation
backgrounds. The propagation includes the energy losses for
protons and nuclei due to interactions with the cosmic microwave
background (CMB) and the extragalactic background light
(EBL)~\cite{bib:rachen}. Two cases were considered: pure proton and pure
iron nuclei emission. The calculations assume an emission power law
spectrum  $dN/dE \propto E^{-\beta}$, with $\beta = 2.4$, $E_{min} =
10^{19.6}$ \eV and  $E_{max} = Z \times10^{21}$ \eV, where $Z$ is the
charge of the cosmic rays.

\begin{figure}[!h]
  \centering
  \subfloat[Swift BAT-70]{\includegraphics[angle=90,width=0.49\textwidth]{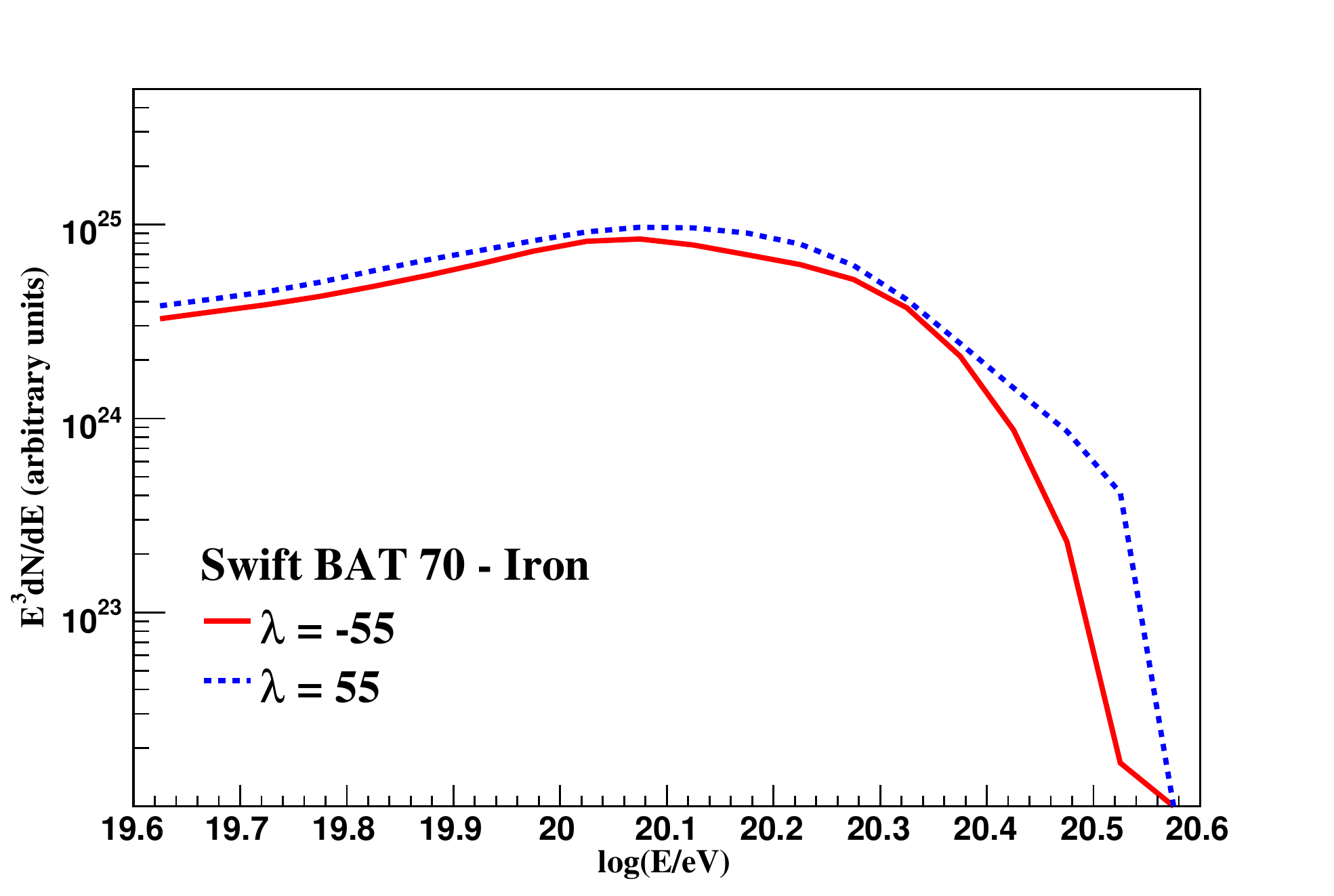}}
  \subfloat[Swift BAT-70 complete]{\includegraphics[angle=90,width=0.49\textwidth]{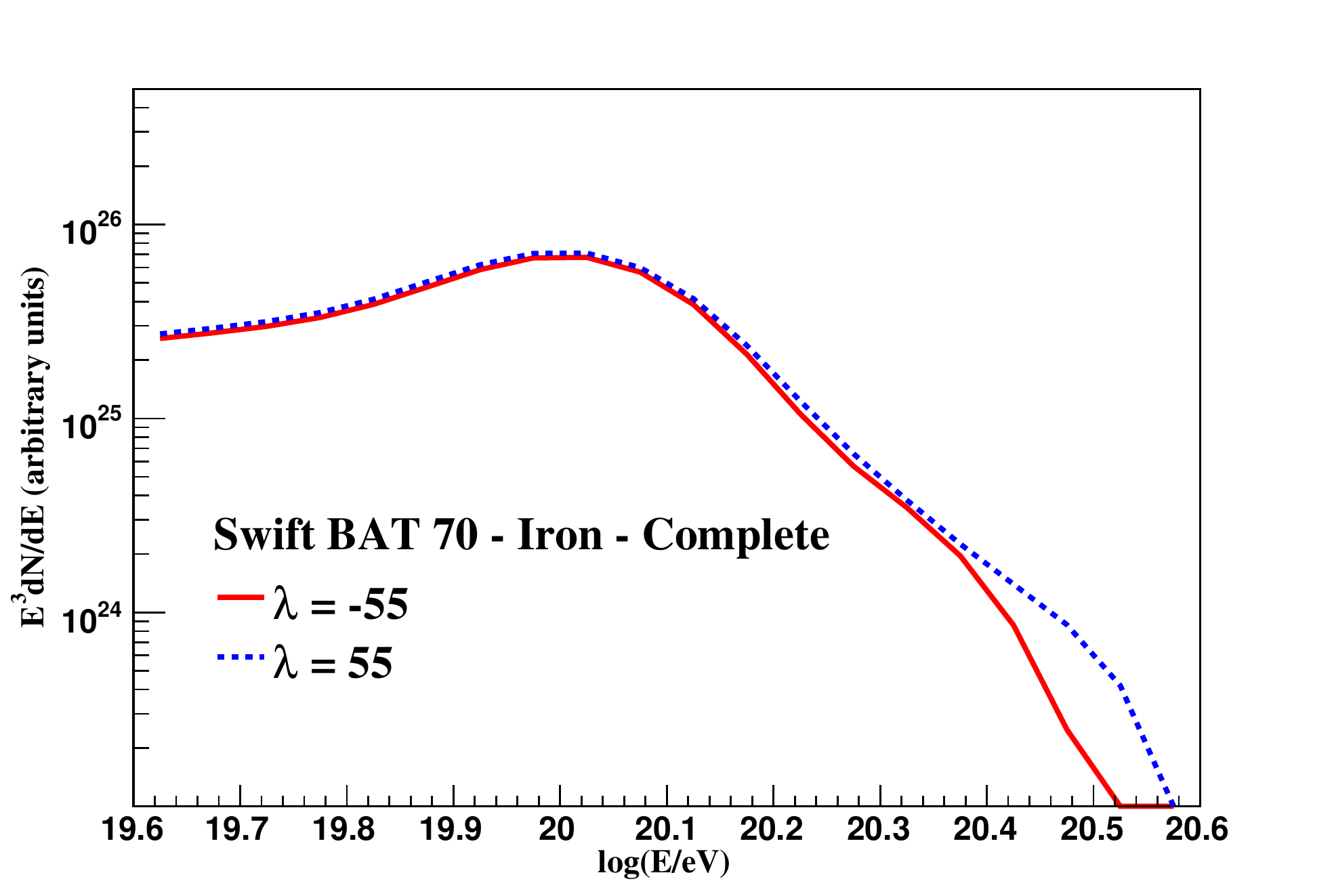}}
  \caption{Energy spectrum measured at Earth as a function of
    energy. Sources from the Swift-BAT catalog (a) and completed
    Swift-BAT (b) with equal UHECR luminosity and $z < 0.072$ were considered. Particles were propagated from source to Earth using the CRPropa program. The contribution of each source was weighted by its exposure as calculated for observatories at $\pm 55$ degrees of latitude. Blue line dashed correspond to $+55$ degrees and red line correspond to $-55$ degrees locations. The simulated spectrum at the source were power law with index $\beta = 2.4$, $E_{min} = 10^{19.6}$ eV and $E_{max} = Z\times10^{21}$ eV.}
  \label{fig:spectrum:lat}
\end{figure}

Particles arriving at Earth from each source were weighted by the
corresponding exposure (equation~\ref{eq:peso}), taking also into account the deflections due to magnetic fields during the particle propagation. Figure~\ref{fig:spectrum:lat} shows as an example the resulting energy spectra for latitudes $\pm 55^\circ$ for pure iron nuclei emitted for sources distributed according to the incomplete (Figure~\ref{fig:spectrum:lat}a)  and completed (Figure~\ref{fig:spectrum:lat}b) Swift-BAT catalog. The effect of the latitude on the energy spectrum is small but clear.

Figures~\ref{fig:spec:porc} and ~\ref{fig:spec:porc:complete} show the percent difference of the flux measured in each hemisphere as a function of energy for latitude $\pm$ 55 degrees for proton and iron leaving the sources. The 2MASS Redshift Survey (2MRS), IRAS 1.2 Jy Survey, Palermo Swift-BAT and Swift-BAT incomplete (Figures~\ref{fig:spec:porc}) and completed (Figure~\ref{fig:spec:porc:complete}) catalogs are shown. For energies above $10^{20.1}$ eV, the number of simulated events arriving on Earth is very small and therefore the results are less statistically significant.

\section{Effect of the latitude on the capability to measure an anisotropy signal}
\label{sec:anisotropy}

Anisotropy of UHECR arrival directions is usually quantified by comparison between the arrival directions of these particles with a template of sources detected in other wavelengths and by independent techniques. Since the exposure of the sources in the catalog is a function of the Observatory latitude, the capability of the method in detecting an anisotropy signal is also a function of latitude.

The standard 2pt method~\cite{bib:2pt:1,bib:2pt:2} was used to quantify
the detection power or detection efficiency of an observatory at a given
latitude. In this method, any departure from isotropy is measured
through a pseudo-log-likelihood function \mbox{$\sigma_P = \sum_i
  \textrm{ln} P_i(n_{obs}|n_{exp})$} in which $P_i$ is the Poisson
distribution: if $k=n_{obs}$, then $P_i(n_{obs}|n_{exp}) = n_{exp}^{k}
\exp{(-n_{exp})}/k!$, $n_{obs}$ is the number of counts observed
and $n_{exp}$ is the expected number of counts from isotropic
samples. The 99\% C.L. significance level for rejecting isotropy was
chosen as a reference. The isotropy expectation at 99\% C.L. significance level ($\sigma_P^{1\%}$) was calculated using $10^5$ events.

Mock skies were generated following the source distribution given by all
the catalogs listed above. Each mock sky was constructed with a limited
number (varying from 10 to 90 in steps of 10) of events allowing us to
study the dependence of the detection power on the number of events
measured by the experiments. The direction of the events were randomly
drawn from the direction of the sources in the catalog listed
above. Therefore each mock sky represents a possible realization of a
limited number of events coming from a given distribution of
sources. The events were weighted according to Equation~\ref{eq:peso},
with exposure map $W_s$ and smeared by a 2D Gaussian 4 degrees wide (standard deviation) to
take into account the possible deflections of the particles, due to the random component of
the magnetic field.

\begin{figure}[!h]
\begin{center}
	\includegraphics[angle=0,width=0.49\textwidth]{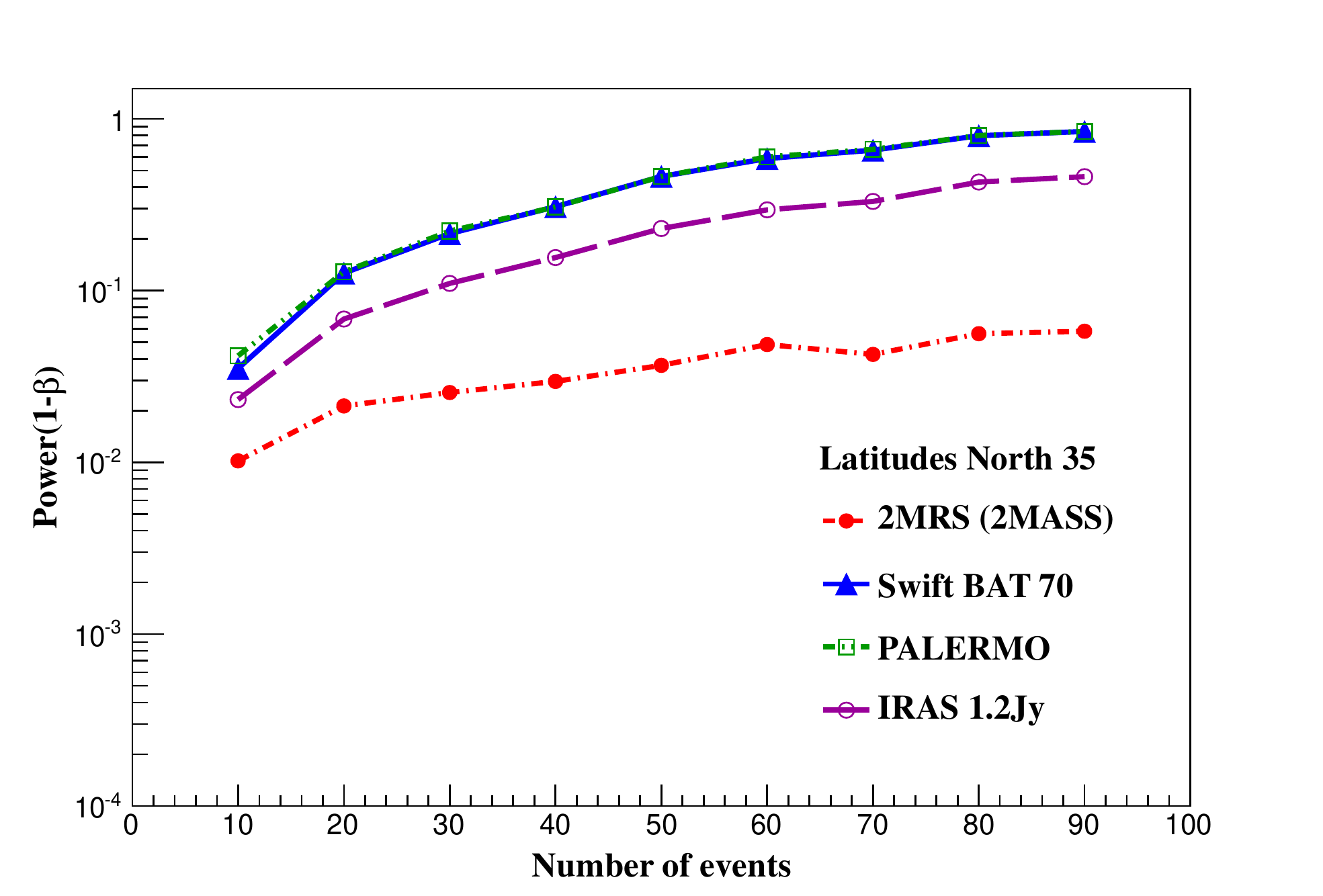}
	\includegraphics[angle=0,width=0.49\textwidth]{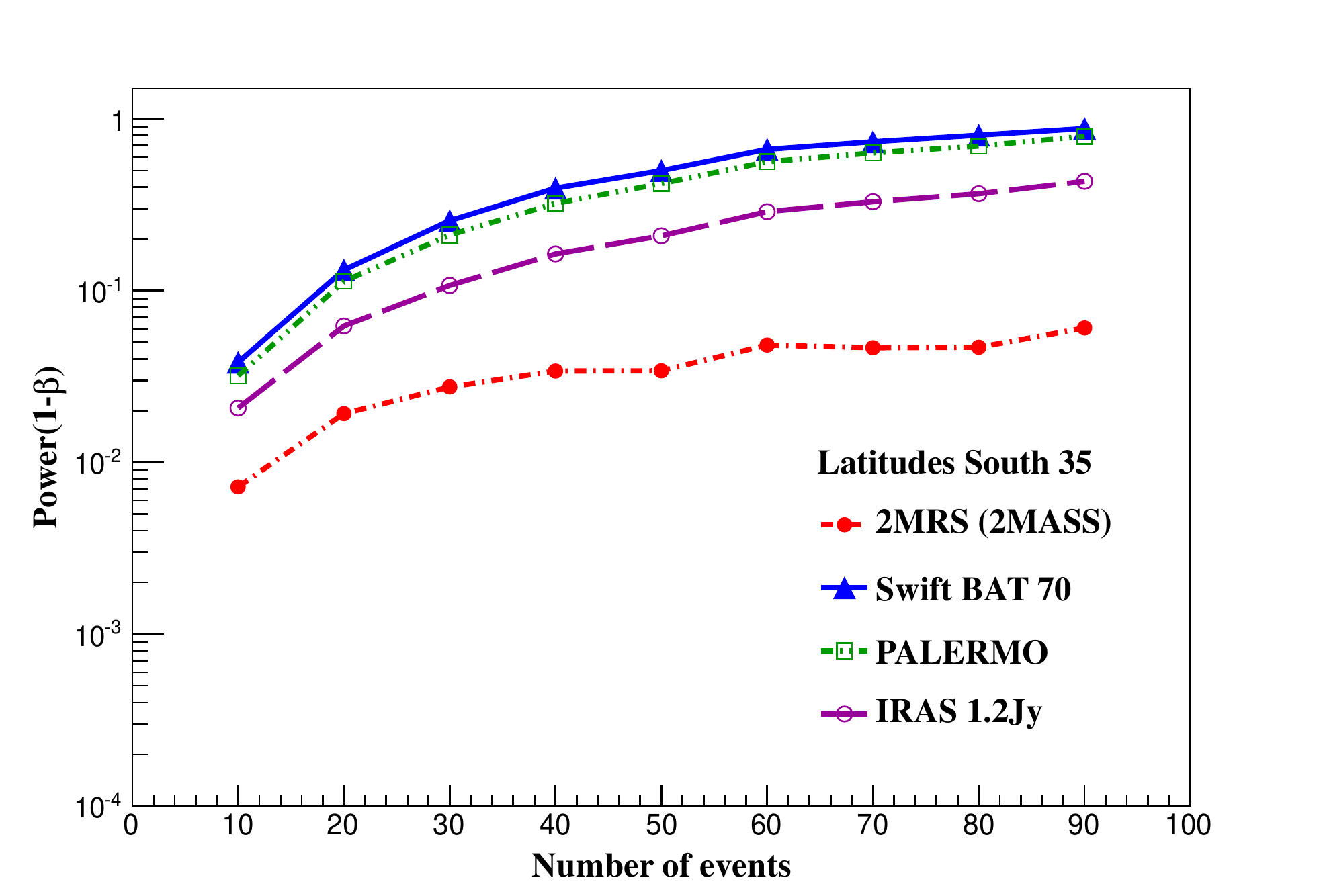}
        \caption{Comparison between the values of power $(1-\beta)$
          according to the number of events for the catalogs 2MASS
          Redshift Survey (2MRS), IRAS 1.2 Jy Survey, Palermo
          Swift-BAT and Swift-BAT to the latitudes $\pm$ 35 and $z < 0.048$ to the Southern and Northern latitudes. The power of the observatory ($1 - \beta$) is defined as the effectiveness of detecting the signal hypothesis. See text for details.}
        \label{fig:power:0048}
\end{center}
\end{figure}

\begin{figure}[!h]
\begin{center}
	\includegraphics[angle=90,width=0.49\textwidth]{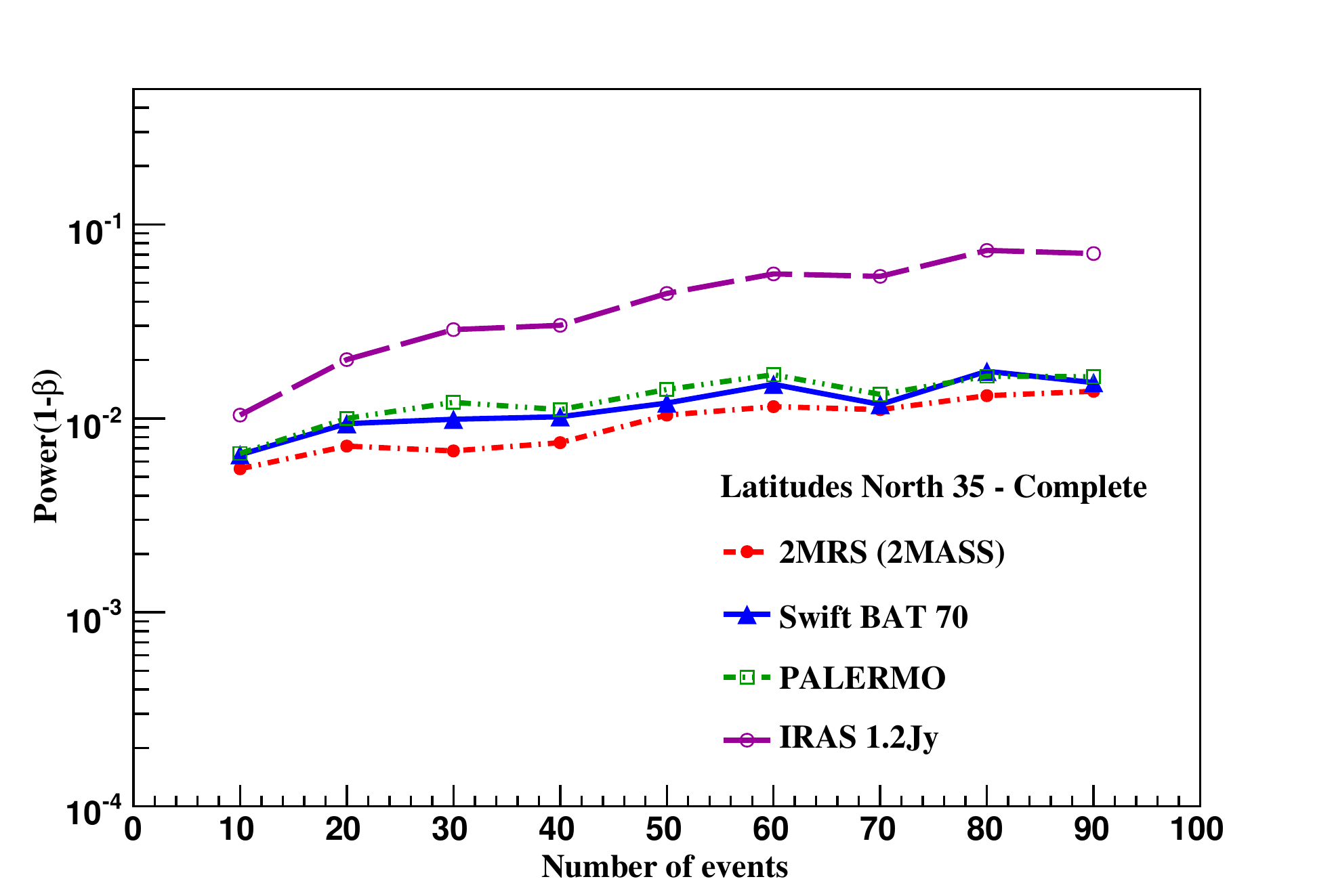}
	\includegraphics[angle=90,width=0.49\textwidth]{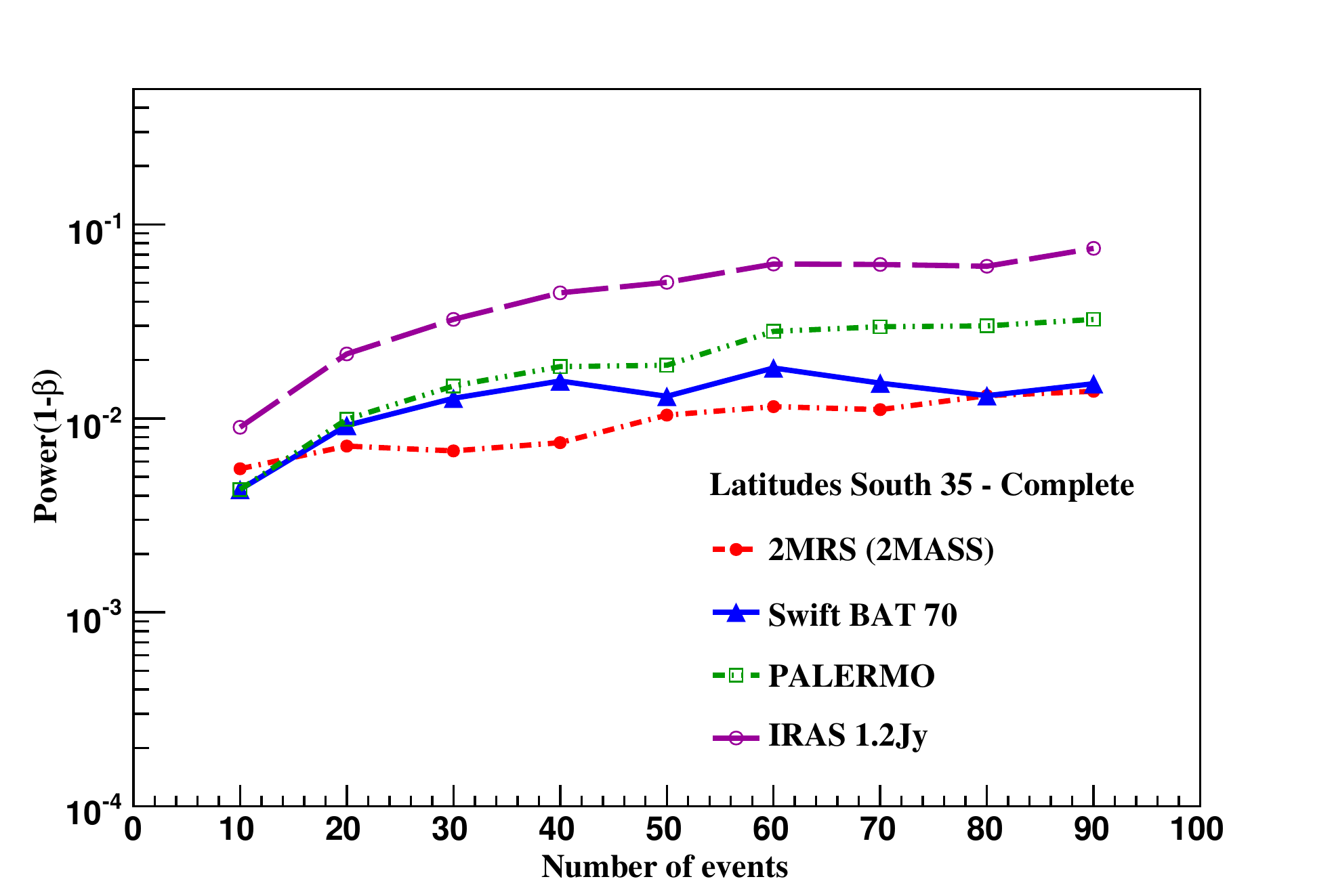}
        \caption{Comparison between the values of power $(1-\beta)$ according to the number of events for the catalogs 2MASS Redshift Survey (2MRS), IRAS 1.2 Jy Survey, Palermo Swift-BAT and Swift-BAT complete to the latitudes $\pm$ 35 and $z < 0.048$ to the Southern and Northern latitudes. The power of the observatory ($1 - \beta$) is defined as the effectiveness of detecting the signal hypothesis. See text for details.}
        \label{fig:power:0048:complete}
\end{center}
\end{figure}

\begin{figure}[!h]
  \begin{center}
    \includegraphics[angle=90,width=0.49\textwidth]{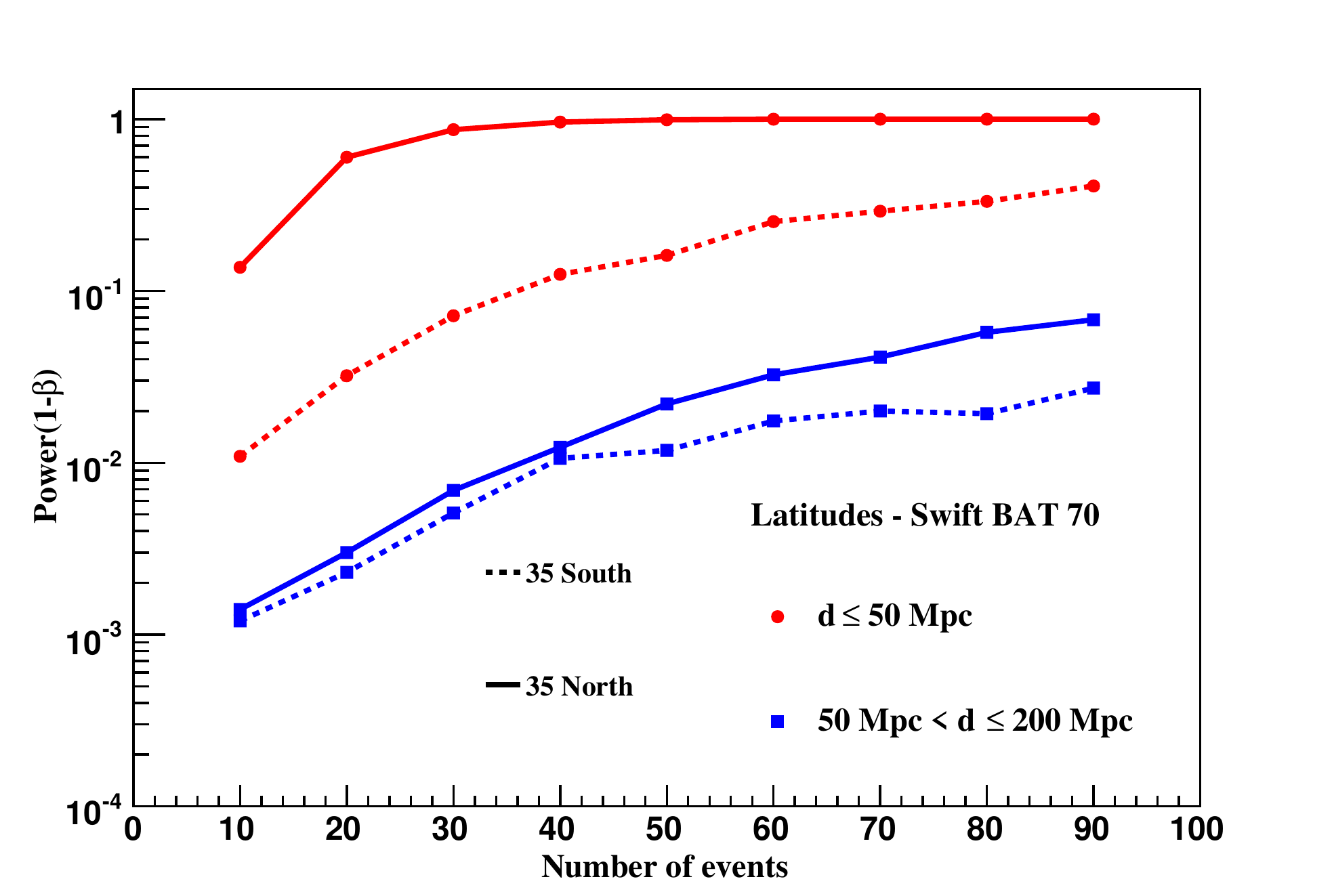}
    \includegraphics[angle=90,width=0.49\textwidth]{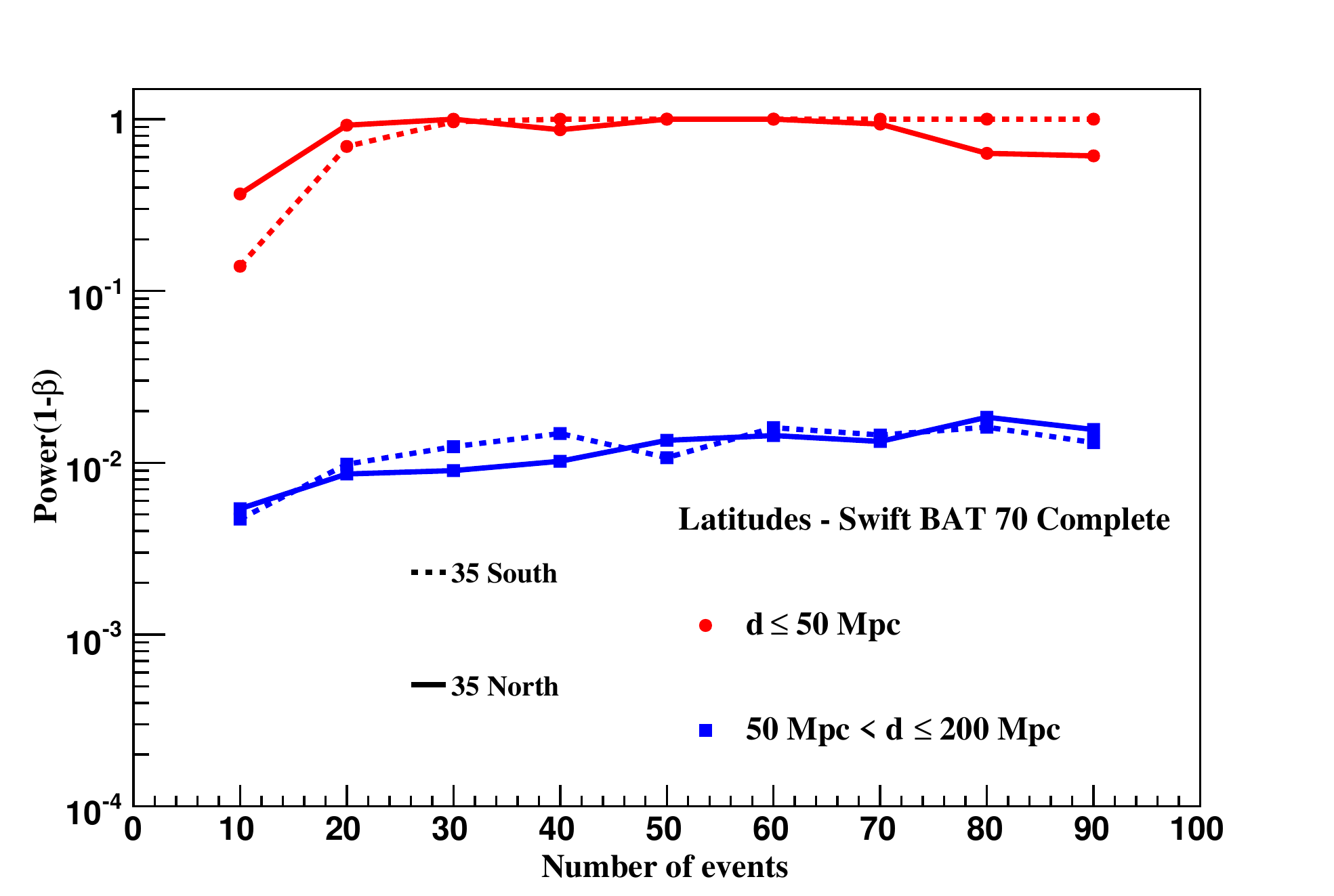}
    \caption{Comparison between the values of power $(1-\beta)$ according to the number of events for the catalog Swift BAT 70 for observatories at latitudes $\pm$ 35 degrees. Red represents sources closer than 50 Mpc and blue represents sources distant 50 to 200 Mpc from Earth. The power of the Observatory ($1 - \beta$) is defined as the effectiveness of detecting the signal hypothesis. See text for details. Left: incomplete catalogue. Right: catalogue completed.}
    \label{fig:anisotropy}
  \end{center}
\end{figure}

The total number of mock skies generated was $10^4$ with the same
number of events, for each mock sky the probability of departure from isotropy was calculated
($\sigma_P$). The probability of measuring anisotropy, given an
anisotropic sky ($\beta$), is given by the ratio of mock skies with
$\sigma_P < \sigma_P^{1\%}$. The detection power of the observatory ($1
- \beta$) is defined as the effectiveness of detecting the signal
hypothesis. Figures~\ref{fig:power:0048} and
\ref{fig:power:0048:complete} show the detection power of observatories
located at $\pm$ 35 degrees as a function of the number of events. The plot on the
left of Figure~\ref{fig:anisotropy} shows that the observatories in the north have a larger power to detect
anisotropy if an anisotropic sky is given. The detection power as a function of the number of events from the catalog Swift-BAT (incomplete and complete) with observatories at latitudes $\pm$ 35 degrees is shown for sources closer (red) and farther (blue) than 50 Mpc. It is clear that the major contribution to the difference between north and south observatories is due to the nearby sources.

We have done the calculations for latitude $\pm$ 25, $\pm$ 35, $\pm$ 45
and $\pm$ 55 degrees. The effect of the latitude on the power to detect
anisotropy using the 2-pt method is important only for latitudes larger
than $\pm$ 35 degrees.

We have simulated sources up to 300 Mpc. The contribution of sources beyond this distance is $\sim$10\% of the total flux. The inclusion of extra isotropic sources beyond this distance would result in a decrease of the detection power calculated in this section for observer in both hemispheres. However if for $D > 300$ Mpc the sources are isotropic for the North and South hemisphere, the conclusions would remain the same. Figures~\ref{fig:power:0048} and~\ref{fig:power:0048:complete} shows differences from North and South hemispheres therefore they would only be affected if sources beyond 300 Mpc were anisotropic in the North and South skies.

\section{Conclusion}
\label{sec:conclusion}

In the present paper, the influence of the latitude of the Observatory
on the measured flux and capability to measure anisotropy of UHECR was
studied. Particles are propagated to Earth from sources distributed
according to the 2MASS Redshift Survey (2MRS), IRAS 1.2 Jy Survey,
Palermo Swift-BAT and Swift-BAT catalogs and the particles were
propagated to Earth. The catalogs were completed with sources
isotropically distributed in the sky and whose distances are such that
the catalog becomes complete above 100 Mpc. The incompleteness of the
catalogs are not important for the calculations because they would
affect equaly observers in both hemispheres. This study considers that sources have the same UHECR intrinsic luminosity. The exposures of observatories at different latitudes were taken into account in order to build the energy spectrum and assess their capability to detect anisotropies.

\begin{figure}[!h]
\begin{center}
	\includegraphics[angle=90,width=0.49\textwidth]{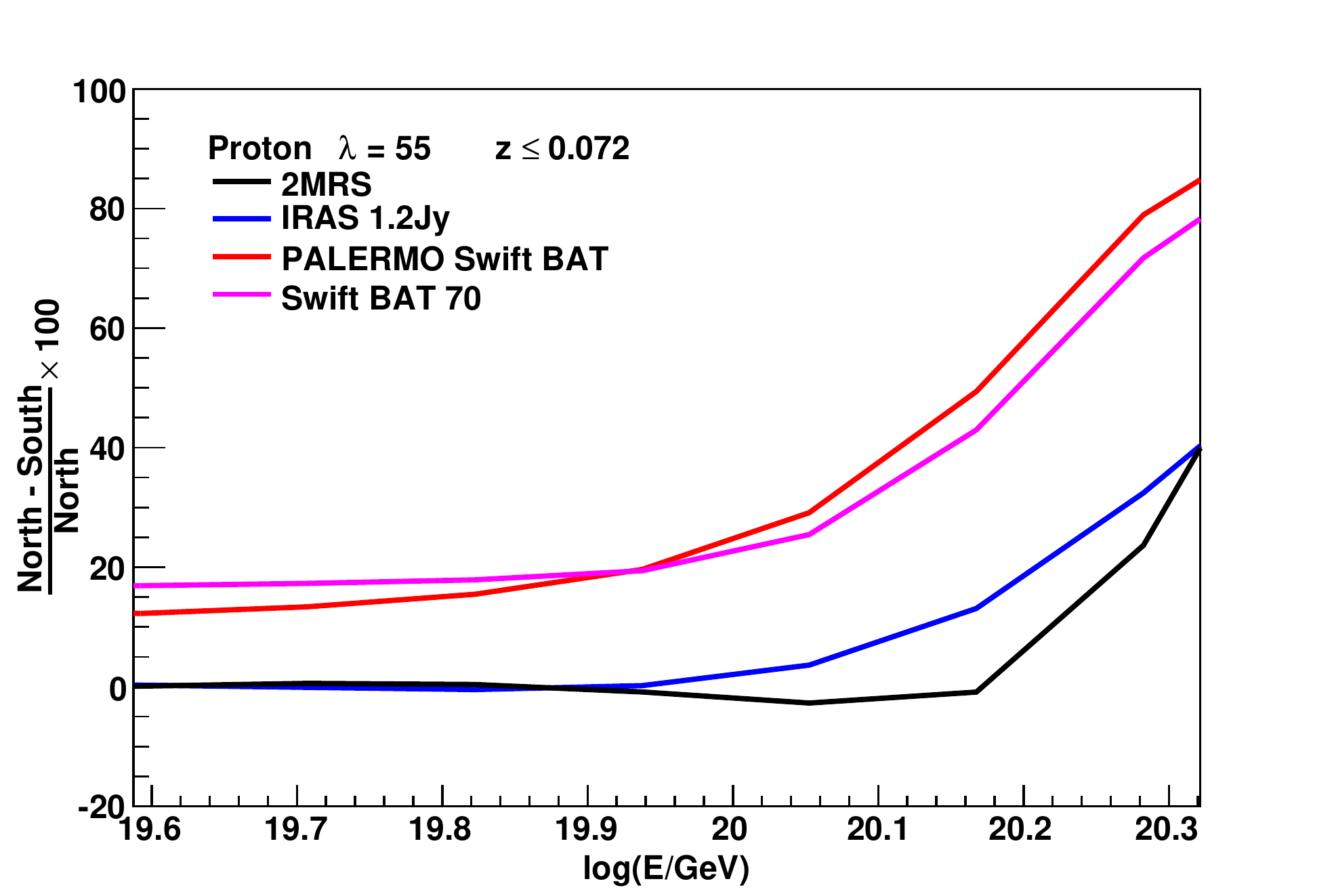}
	\includegraphics[angle=90,width=0.49\textwidth]{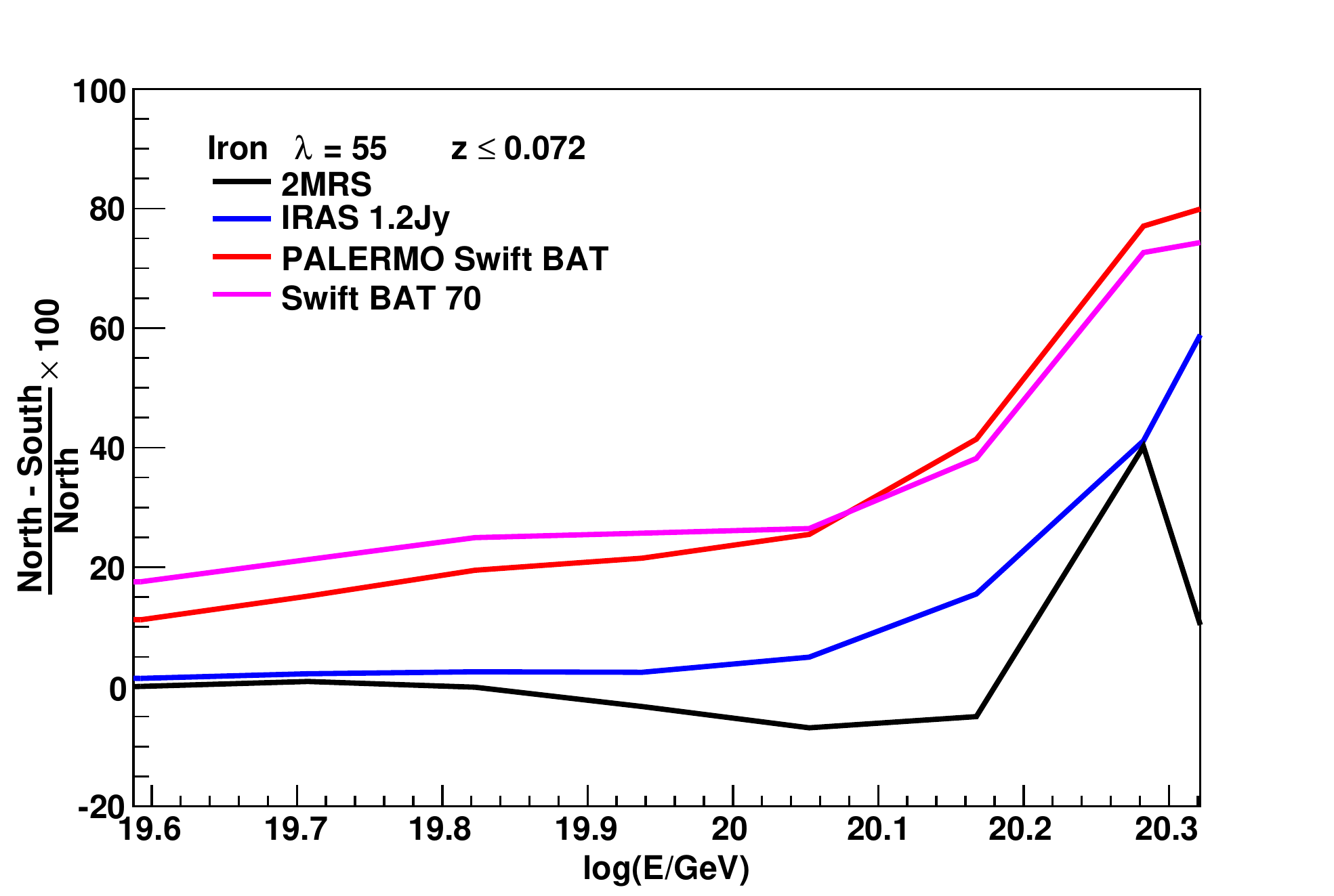}
        \caption{Relative difference of the flux measured by observatories located in the northern and southern hemispheres at equal latitudes. Sources from the catalogs with equal UHECR luminosity and $z < 0.072$ were considered. Particles were propagated from source to Earth using the CRPropa program. The contribution of each source was weighted by its exposure. In the left panel is shown the case in which only proton have been emitted by the sources. In the right panel is shown the case in which only iron nuclei have been emitted by the sources. The simulated spectra at the source were power laws with index $\beta = 2.4$, $E_{min} = 10^{19.6}$ eV and $E_{max} = Z\times10^{21}$ eV.}
        \label{fig:spec:porc}
\end{center}
\end{figure}

\begin{figure}[!h]
\begin{center}
	\includegraphics[angle=90,width=0.49\textwidth]{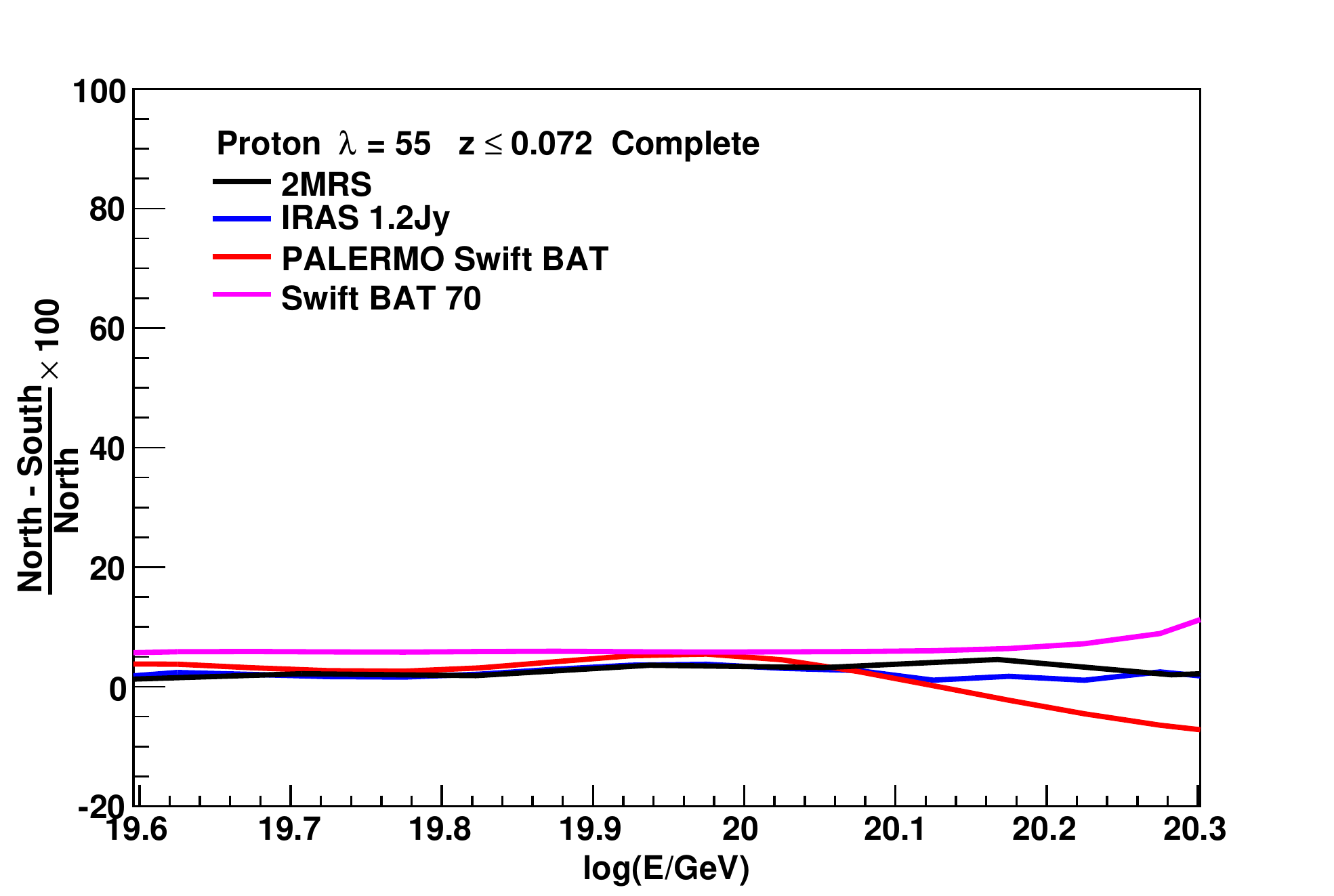}
	\includegraphics[angle=90,width=0.49\textwidth]{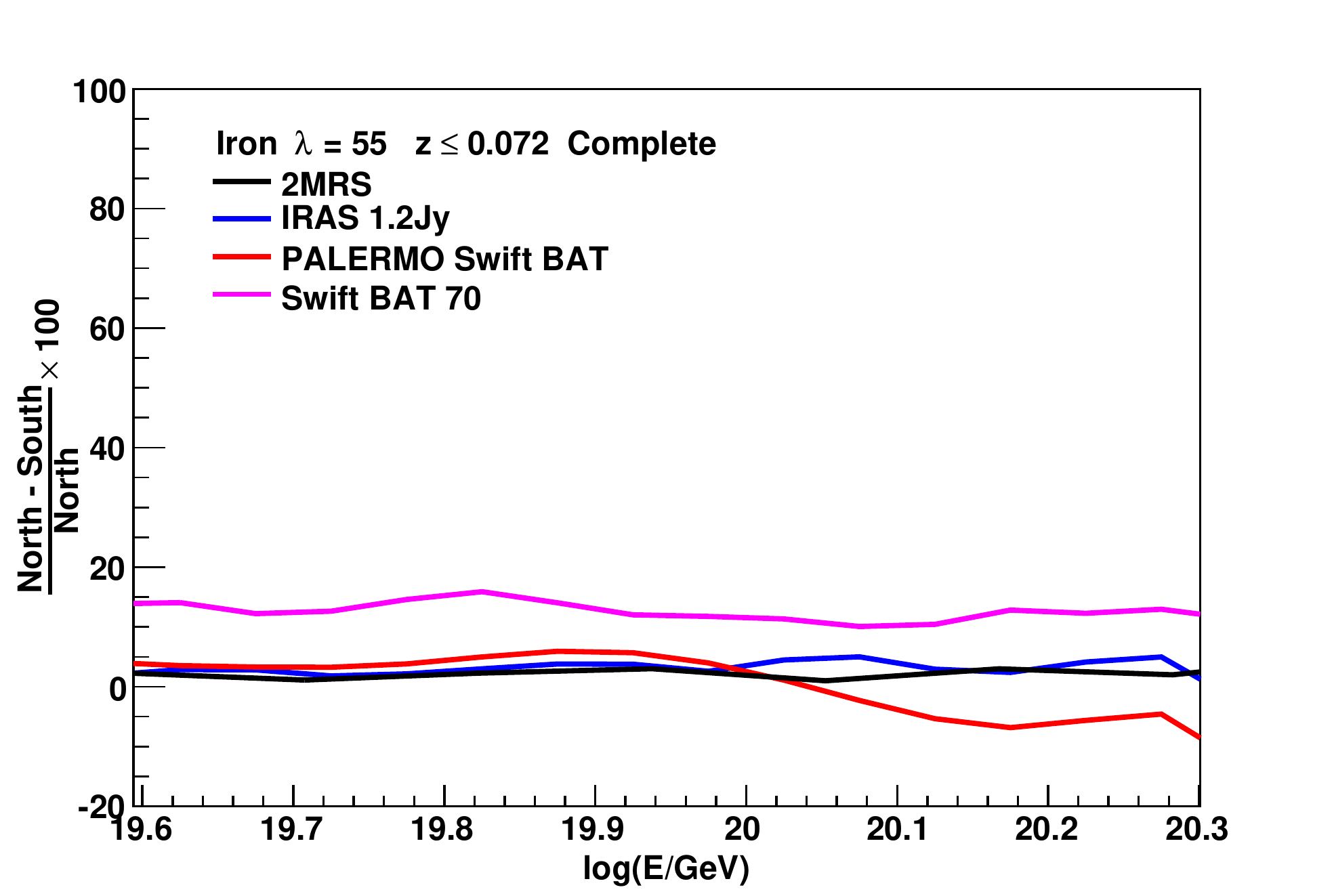}
        \caption{Relative difference of the flux measured by
          observatories located in the northern and southern hemispheres
          at equal latitudes. Sources from the completed catalogs with equal UHECR luminosity and $z < 0.072$ were considered. Particles were propagated from source to Earth using the CRPropa program. The contribution of each source was weighted by its exposure. In the left panel is shown the case in which only proton have been emitted by the sources. In the right panel is shown the case in which only iron nuclei have been emitted by the sources. The simulated spectra at the source were power laws with index $\beta = 2.4$, $E_{min} = 10^{19.6}$ eV and $E_{max} = Z\times10^{21}$ eV.}
        \label{fig:spec:porc:complete}
\end{center}
\end{figure}

The influence of the latitude on the flux was quantified as a function of energy (see Figures~\ref{fig:spec:porc} and ~\ref{fig:spec:porc:complete}). The differences between the flux measured by north and south observatories at $\pm$ 55 degrees for energies above $10^{19.6}$ eV can be as large as 20\% for incomplete catalogs and 5\% for completed catalogs. The influence of the latitude on the power to detect anisotropy, by comparing mock skies generated according to the sources distribution from the catalogs, was also calculated. Observatories in the northern hemisphere with latitude larger than 35 degrees have a greater capability to determine an anisotropic sky than the southern ones for the catalogs used here as anisotropy templates for anisotropy.

The calculations done here show that an anisotropy of the sources of UHECR breaks the symmetry between northern and southern observatories and introduces a dependence of the total measured flux for latitudes larger than $\pm$ 45 degrees and of the capability to determine anisotropic skies on the latitude of the observatory for latitudes larger than $\pm$ 35 degrees. The effect of the magnetic field was taken into account by performing an angular smearing of arrival directions. Therefore, the effect of an anisotropic sky should be taken into consideration when common interpretations of observatories in different latitudes are constructed.

\acknowledgments

This work is funded by FAPESP (2010/07359-6,2014/19946-4), CAPES and CNPq.

\end{document}